\documentclass[twocolumn,aps, superscriptaddress]{revtex4}
\usepackage{graphicx}
\usepackage{amsmath}
\newcommand{\grscale}{0.47}
\begin{document}

\title{Fourier analysis of Ramsey fringes observed in a continuous \\ 
atomic fountain for in situ magnetometry}

\author{Gianni Di Domenico}
\affiliation{Laboratoire Temps-Fr\'equence, Universit\'e de Neuch\^atel, \\
Avenue de Bellevaux 51, CH-2000 Neuch\^atel, Switzerland}
\author{Laurent Devenoges}
\affiliation{Laboratoire Temps-Fr\'equence, Universit\'e de Neuch\^atel, \\
Avenue de Bellevaux 51, CH-2000 Neuch\^atel, Switzerland}
\author{Andr\'e Stefanov}
\affiliation{Swiss federal office of metrology, METAS\\
Lindenweg 50, CH-3003 Bern-Wabern, Switzerland}
\author{Alain Joyet}
\affiliation{Laboratoire Temps-Fr\'equence, Universit\'e de Neuch\^atel, \\
Avenue de Bellevaux 51, CH-2000 Neuch\^atel, Switzerland}
\author{Pierre Thomann}   
\affiliation{Laboratoire Temps-Fr\'equence, Universit\'e de Neuch\^atel, \\
Avenue de Bellevaux 51, CH-2000 Neuch\^atel, Switzerland}

%
%\date{Received: 01.04.2011 / Revised version: 01.04.2011}

\begin{abstract}
Ramsey fringes observed in an atomic fountain are formed by the superposition of the individual atomic signals. Due to the atomic beam residual temperature, the atoms have slightly different trajectories and thus are exposed to a different average magnetic field, and a velocity dependent Ramsey interaction time. As a consequence, both the velocity distribution and magnetic field profile are imprinted in the Ramsey fringes observed on Zeeman sensitive microwave transitions. In this work, we perform a Fourier analysis of the measured Ramsey signals to retrieve both the time averaged magnetic field associated with different trajectories and the velocity distribution of the atomic beam. We use this information to reconstruct Ramsey fringes and establish an analytical expression for the position of the overall observed Ramsey pattern.
\end{abstract}
\maketitle

%%%%%%%%%%%%%%%%%%%%%%%%%%%%%%%%%%%%%%%%%%%%%%%%%%%%%%%%%%%%%%%%%%%%%%%%%%%%%
\section{Introduction}
\label{section0}
%%%%%%%%%%%%%%%%%%%%%%%%%%%%%%%%%%%%%%%%%%%%%%%%%%%%%%%%%%%%%%%%%%%%%%%%%%%%%

Since its invention in 1950, the method of separated oscillatory fields~\cite{Ramsey1950} has been widely used in physics experiments. Particularly in the field of atomic clocks where it made possible successive improvements of the performances by many orders of magnitude. Indeed, by exciting the atomic transition with separated oscillatory fields, the width of the resonance is inversely proportional to the free evolution time between the two excitation pulses. As a consequence, any method allowing to increase the free evolution time results in an improvement of the clock performance.

The separated oscillatory fields method was first applied to thermal atomic beams where the free evolution time is of the order of $10^{-3}$-$10^{-2}$~s, limited by the length of the resonator. With the advent of laser cooling, it became possible to produce fountains of cold atoms~\cite{Clairon1991}, and thereby to increase the free evolution time to approximately $0.5$~s mainly limited by the free fall due to the earth gravitational field and geometrical constraints. All atomic fountain clocks using laser cooled atoms that currently contribute to TAI (International Atomic Time) are based on a pulsed mode of operation: atoms are sequentially laser-cooled, launched vertically upwards and interrogated during their ballistic flight before the cycle starts over again~\cite{Wynands2005}. This approach has made possible important advances in time and frequency metrology: state-of-the-art fountains are operated in National metrological institutes at an accuracy level below $10^{-15}$ in relative units. 

Our alternative approach to atomic fountain clocks is based on a continuous beam of laser-cooled atoms. Besides making the intermodulation effect negligible~\cite{Joyet2001,GuenaPaper}, a continuous beam is also interesting from the metrological point of view. Indeed, the relative importance of the contributions to the error budget is different for a continuous fountain than for a pulsed one, notably for density related effects (collisional shift), which are an important issue if high stability and high accuracy are to be achieved simultaneously.

As a motivation for our work, evaluation of the second order Zeeman shift in atomic fountain clocks requires a precise knowledge of the magnetic field in the free evolution zone. The methods developed in pulsed fountains to map the magnetic field in the resonator are based on throwing balls of atoms at different altitudes. This is not applicable to our continuous fountain since the atomic trajectory is not vertical and therefore the launching velocity range is limited by geometrical constraints. Moreover, the atomic beam longitudinal temperature is higher in our continuous fountain ($75$~$\mu$K) than in pulsed fountains ($1$~$\mu$K) and as a consequence the distribution of apogees is wider. The effect of this large distribution of transit time is to modify significantly the Ramsey pattern, reducing the number of fringes but also increasing its dependance on magnetic field inhomogeneities. 

The work presented in this article is devoted to developping a new method to investigate the magnetic field in the atomic resonator where the free evolution takes place. In thermal beams standards, the shape of the Ramsey signal has been used as a diagnostic tool to measure the distribution of transit times and thus the atomic beam longitudinal velocity distribution~\cite{Jarvis74}, \cite{Boulanger86}, \cite{Makdissi97}. In this article we will show that, in a continuous atomic  fountain, the Fourier analysis of Zeeman sensitive Ramsey fringes allows one to measure the time-averaged magnetic field seen by the atoms during their free evolution. Moreover, it gives a better understanding of the shape of the Ramsey pattern that we observe in our continuous atomic fountain. More precisely, it helps one to understand the difference between the position of the central fringe (for which the microwave is in phase with the atomic dipole) and the position of the fringe which shows the highest contrast.

In section~\ref{section1} we will give a brief description of our continuous atomic fountain FOCS-2. Then we will explain the principle of our analysis in section~\ref{section2} and present the experimental procedure and results in section~\ref{section4}. Finally we will discuss and interpret the experimental results in section~\ref{section5} and conclude in section~\ref{section6}.

%%%%%%%%%%%%%%%%%%%%%%%%%%%%%%%%%%%%%%%%%%%%%%%%%%%%%%%%%%%%%%%%%%%%%%%%%%%%%
\section{Continuous atomic fountain clock FOCS-2}
\label{section1}
%%%%%%%%%%%%%%%%%%%%%%%%%%%%%%%%%%%%%%%%%%%%%%%%%%%%%%%%%%%%%%%%%%%%%%%%%%%%%

\begin{figure}[tbp]
\includegraphics[width=\grscale\textwidth]{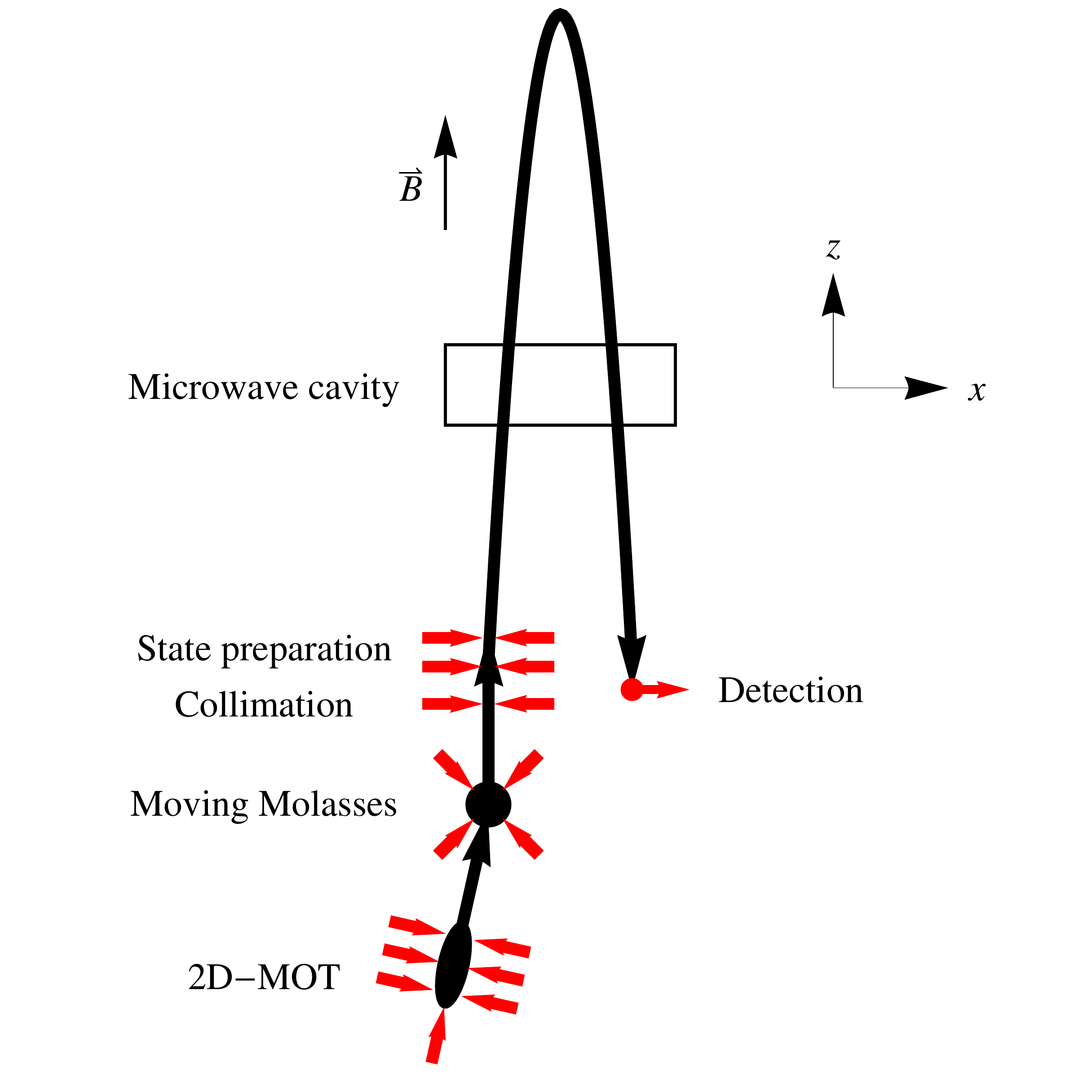}
\caption{Scheme of the continuous atomic fountain clock FOCS-2. An intense atomic beam of pre-cooled cesium atoms is produced in the two-dimensional magneto-optical trap (2D-MOT). The atoms are then captured by the 3D moving molasses which further cools and launches the atoms at a speed of $4$~m/s. Then the atomic beam is collimated with Sisyphus cooling in the transverse directions, and before entering the microwave cavity, the atoms are pumped in $\left|F=3,m=0\right\rangle$ by state preparation. Finally, after the second passage in the microwave cavity, the transition probability is measured by fluorescence detection of the atoms in $F=4$. The transit time between the two $\pi/2$-pulses is $T\approx 0.5$~s.}
\label{fig:Fig0}
\end{figure}

In our experiment, we use the separated oscillatory fields method~\cite{Ramsey1950} to measure the transition probability of cesium atoms between $\left|F=3,m_\mathrm{F}\right\rangle$ and $\left|F=4,m_\mathrm{F}\right\rangle$ for $m_\mathrm{F}=-3,\cdots,3$. A scheme of the continuous atomic fountain clock FOCS-2 is shown in Fig.~\ref{fig:Fig0}. The atomic beam is produced with a two-dimensional magneto-optical trap~\cite{CastagnaPaper}. The atoms are further cooled and launched at a speed of $4$~m/s with the moving molasses technique~\cite{PBe99}. The longitudinal temperature at the exit of the moving molasses is $75$~$\mu$K. Before entering the microwave cavity, the atomic beam is collimated by transverse Sisyphus cooling and the atoms are pumped into $F=3$ with a state preparation scheme combining optical pumping with laser cooling~\cite{DiDomenicoPRA2010}. After these two steps, the transverse temperature is decreased to approximately $3$~$\mu$K. During the first passage into the microwave cavity, we apply a $\pi/2$-pulse (duration of $10$~ms) and thereby produce a superposition state which evolves freely for approximately $0.5$~s before the second $\pi/2$-pulse is applied during the second passage. Finally, the transition probability between $\left|F=3,m_\mathrm{F}\right\rangle$ and $\left|F=4,m_\mathrm{F}\right\rangle$ is measured by fluorescence detection of the atoms in $F=4$.

Ramsey fringes are obtained by measuring the transition probability as a function of the microwave frequency, which is scanned around each of the hyperfine transitions between the states $\left|F=3,m_\mathrm{F}\right\rangle$ and $\left|F=4,m_\mathrm{F}\right\rangle$. A magnetic field is used in the interogation zone to lift the degeneracy of the Zeeman sub-levels. The Ramsey fringes are formed by the superposition of individual atomic signals. Because of the residual atomic beam temperature in the longitudinal direction, every atom has a slightly different trajectory with a different transit time. Moreover, the magnetic field in the free evolution zone has small inevitable spatial variations and therefore the average magnetic field seen by the atoms depends on the trajectory. As a consequence, information on both the atomic velocity distribution and the magnetic field profile is contained in the experimental Ramsey fringes for $m_\mathrm{F}\neq 0$. Our objective is to retrieve this information from a measurement of Ramsey fringes.

%%%%%%%%%%%%%%%%%%%%%%%%%%%%%%%%%%%%%%%%%%%%%%%%%%%%%%%%%%%%%%%%%%%%%%%%%%%%%
\section{Principle}
\label{section2}
%%%%%%%%%%%%%%%%%%%%%%%%%%%%%%%%%%%%%%%%%%%%%%%%%%%%%%%%%%%%%%%%%%%%%%%%%%%%%

As explained in the previous section, the Ramsey fringes measured in our continuous atomic fountain are formed by the superposition of Ramsey signals coming from individual atoms. The contribution of an individual atom to the Ramsey fringes observed on the Zeeman component $m_\mathrm{F}$ is given by:
\begin{equation}
	I_{m_\mathrm{F}}(\omega_\mathrm{rf},T)=\frac{1}{2}I_0\sin^2(b\,\tau)\left[ 1+\cos\left( \varphi_{m_\mathrm{F}}(\omega_\mathrm{rf},T) \right)\right]
\end{equation}
where $b$ is the Rabi angular frequency, $\tau$ is the microwave interaction time, $\omega_\mathrm{rf}$ is the microwave frequency, $T$ is the effective transit time between the first and second $\pi/2$-pulses, $I_0$ is a global amplitude factor, and the phase $\varphi_{m_\mathrm{F}}(\omega_\mathrm{rf},T)$ is given by :
\begin{equation}
	\varphi_{m_\mathrm{F}}(\omega_\mathrm{rf},T)=\int_0^T \left[ \omega_\mathrm{rf}-\omega_{0}-m_\mathrm{F} 2\pi K_z B\left(z(t)\right)\right]\,dt
\end{equation}
In this last equation, $\omega_{0}$ is the frequency of the magnetic field-insensitive clock transition, the linear Zeeman shift sensitivity constant is $K_z=7$ Hz/nT, and $B\left(z(t)\right)$ is the magnetic field seen by the atom during its free evolution. The magnetic field in the interrogation region can be separated in a constant value plus residual spatial variations\footnote{In principle, the choice of the constant part $B_0$ is arbitrary and does not influence the present analysis. In practice, we will choose $B_0$ such as to minimize phase variations of the Fourier transform of Ramsey fringes (see section~\ref{section4}) which is equivalent to choose $B_0=\overline{B}(\overline{T})=\frac{1}{\overline{T}}\int_0^{\overline{T}} B\left(z(t)\right)\,dt$ where $\overline{T}$ is the average transit time and $z(t)$ the atomic beam trajectory.}:
\begin{equation}
	B(z)=B_0+B_\mathrm{res}(z)
\end{equation}
Therefore, by introducing the following definitions, firstly for the microwave detuning with respect to the atomic transition:
\begin{equation}
	\Omega=\omega_\mathrm{rf}-\omega_{0}-m_\mathrm{F} 2\pi K_z B_0
\label{defOmega}
\end{equation}
and secondly for the residual phase:
\begin{equation}
	\varphi_\mathrm{res}(T)=2\pi K_z \int_0^T B_\mathrm{res}\left(z(t)\right)\,dt
\label{defphires}
\end{equation}
one can write:
\begin{equation}
	\varphi_{m_\mathrm{F}}(\Omega,T)=\Omega\, T - m_\mathrm{F} \varphi_\mathrm{res}(T)
\end{equation}
The total signal is given by adding the contributions from different velocity classes:
\begin{eqnarray}
	I_{m_\mathrm{F}}(\Omega)&=&\frac{I_0}{2}\int_0^{\infty}\hat{\rho}(T)\left[ 1+\cos\left( \varphi_{m_\mathrm{F}}(\Omega,T) \right)\right]\,dT \\
	&=&\frac{I_0}{2}\left[ 1+\mathrm{Re}\int_0^{\infty}\hat{\rho}(T)e^{i\varphi_{m_\mathrm{F}}(\Omega,T)}dT\right] \label{eqmf}\\
	&=&\frac{I_0}{2}+\frac{I_0}{4}\int_{-\infty}^{\infty}\hat{\rho}(T)e^{-i m_\mathrm{F} \varphi_\mathrm{res}(T)}e^{i\Omega\, T}dT\,
	\label{eqItot}
\end{eqnarray}
where $\hat{\rho}(T)=\rho(T)\sin^2(b\,\tau)$, $\rho(T)$ is the transit time distribution, and the last equality is valid if one extends the definition of both $\hat{\rho}(T)$ and $\varphi_\mathrm{res}(T)$ to negative values of the transit time as follows $\hat{\rho}(-T)=\hat{\rho}(T)$ and $\varphi_\mathrm{res}(-T)=-\varphi_\mathrm{res}(T)$. From equation (\ref{eqItot}), one can see that the Fourier transform of $I_{m_\mathrm{F}}(\Omega)$ is given by:
\begin{equation}
	\mathrm{FT}\left\{I_{m_\mathrm{F}}(\Omega)\right\}
	=\frac{I_0}{4}\left[2\delta (T) + \hat{\rho}(T)e^{-i m_\mathrm{F} \varphi_\mathrm{res}(T)}\right]
	\label{FTequ}
\end{equation}
where $\delta$ is the Dirac delta function. In other words, the Fourier transform of the Ramsey signal gives access to the distribution of transit times {\it and} to the dephasing induced by residual magnetic field spatial variations when $m_\mathrm{F}\neq 0$.

%%%%%%%%%%%%%%%%%%%%%%%%%%%%%%%%%%%%%%%%%%%%%%%%%%%%%%%%%%%%%%%%%%%%%%%%%%%%%
\section{Experimental results}
\label{section4}
%%%%%%%%%%%%%%%%%%%%%%%%%%%%%%%%%%%%%%%%%%%%%%%%%%%%%%%%%%%%%%%%%%%%%%%%%%%%%

\begin{figure*}[tbp]
\includegraphics[width=\textwidth]{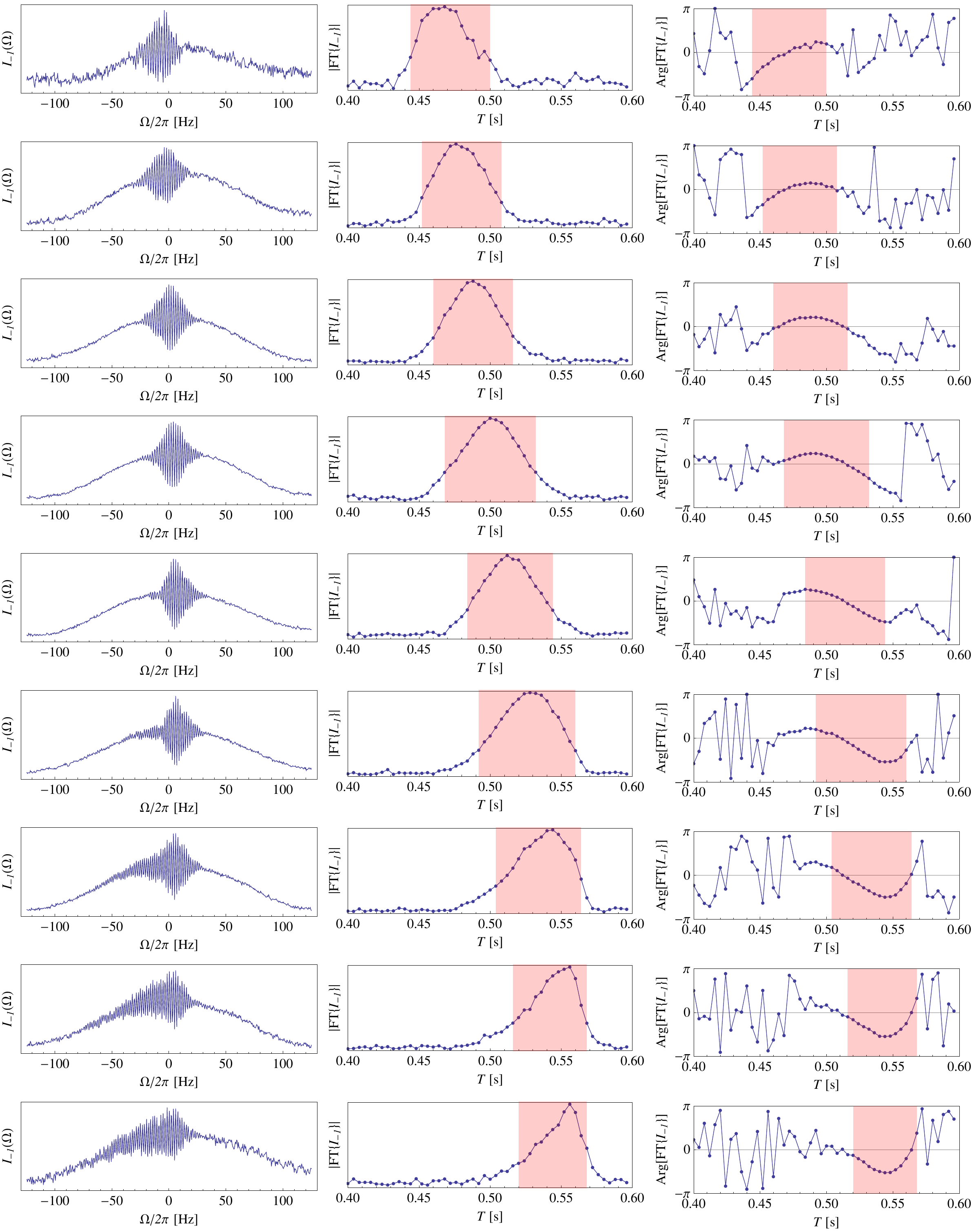}
\caption{The first column shows the Ramsey signals $I_{-1}(\Omega)$ measured on $m_\mathrm{F}=-1$ for various launching velocities of $3.74$~m/s, $3.80$~m/s, $3.86$~m/s, $3.92$~m/s, $3.98$~m/s, $4.04$~m/s, $4.1$~m/s, $4.16$~m/s, $4.22$~m/s from top to bottom. The second and third columns are the module and the phase of the Fourier transform $\mathrm{FT}\left\{I_{-1}(\Omega)\right\}$. They give access, respectively, to the distribution of transit time $\rho(T)$ and to the residual phase change $\varphi_\mathrm{res}(T)$ due to magnetic field spatial variations. Note that the phase is meaningful only in the region where the module is different from zero (highlighted in red). The value of $B_0$ in Eq.~(\ref{defOmega}) is $73.4$~nT, chosen such that the fringe pattern is centered on $\Omega=0$ (or equivalently to minimize the range of the phase).}
\label{fig:Fig1}
\end{figure*}

\begin{figure}[tbp]
\includegraphics[width=\grscale\textwidth]{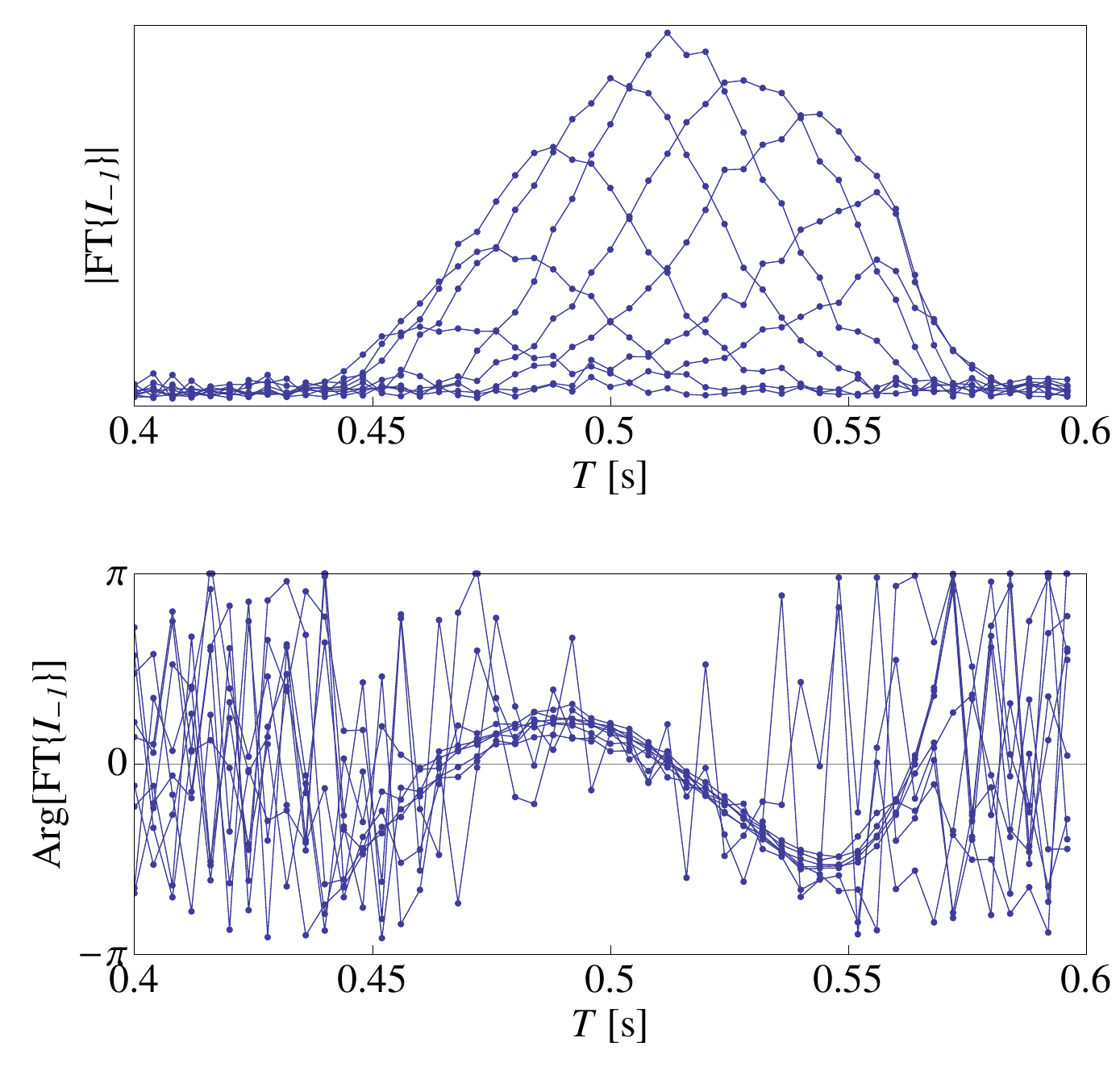}
\caption{Fourier transforms of the Ramsey fringes measured on $m_{\mathrm{F}}=-1$ for different launching velocities (see also Fig.~\ref{fig:Fig1}). The signals corresponding to different velocities are superposed on the same graphs. The upper graph shows the module of the Fourier transforms which is interpreted as a distribution of transit times. The lower graph shows the phase of the Fourier transforms which is interpreted as the dephasing due to magnetic field residual inhomogeneities. See section~\ref{section4} for details.}
\label{fig:Fig2}
\end{figure}

In order to verify the Fourier relation presented in Eq.~(\ref{FTequ}), we measured the Ramsey fringes in our continuous atomic fountain for all $m_\mathrm{F}$ values and for different launching velocities of the moving molasses. Due to geometrical constraints (the atoms have to pass through the two holes of the microwave cavity) the launching velocity can be changed between $3.74$~m/s and $4.22$~m/s. The atomic flux decreases to zero outside of this range, and it is maximum for $4.0$~m/s. The experimental results are displayed in Fig.~\ref{fig:Fig1} for $m_\mathrm{F}=-1$ and in increasing order of the launching velocity ($3.74$~m/s, $3.80$~m/s, $3.86$~m/s, $3.92$~m/s, $3.98$~m/s, $4.04$~m/s, $4.1$~m/s, $4.16$~m/s, $4.22$~m/s). The first column shows the measured Ramsey signals $I_{-1}(\Omega)$ as a function of the microwave detuning $\Omega$. The second and third columns display the module and the phase of the Fourier transform $\mathrm{FT}\left\{I_{-1}(\Omega)\right\}$ respectively. According to Eq.~(\ref{FTequ}):
\begin{itemize}
	\item $\left|\mathrm{FT}\left\{I_{-1}(\Omega)\right\}\right|$, displayed in the second column, is proportional to the distribution of transit time $\rho(T)$. The relation between the Fourier transform of the Ramsey fringes pattern pattern and the velocity distribution has already been studied for thermal beams~\cite{Shirley89}, \cite{Shirley97}. However because of the broad velocity distribution of thermal beams, the interaction time in the cavity cannot be considered as constant, which makes the analysis of the Ramsey fringes more complicated.  On the contrary in FOCS-2, the velocity distribution is sufficiently narrow (see section~\ref{secVT}) such that $\hat{\rho}(T)\approx\rho(T)$ at optimum power, the difference being smaller than $1$\%.
	\item $\mathrm{Arg}\left[\mathrm{FT}\left\{I_{-1}(\Omega)\right\}\right]$, displayed in the third column, is equal to the residual phase change $\varphi_\mathrm{res}(T)$ due to magnetic field spatial variations.
\end{itemize}
In Fig.~\ref{fig:Fig2} we show the same series of measurements but superposed on the same graphs. We observe that the modules of the Fourier transforms are bell-shaped curves whose centre of gravity shifts to the right for increasing launching velocities, in agreement with their interpretation as distribution of transit times. On the other hand, the phases of the Fourier transforms superpose to each other in the domains of $T$ values where the module is different from zero. It is thus possible to measure the residual phase $\varphi_\mathrm{res}(T)$ in the range of $T$ values accessible in the experiment i.e. between $0.44$~s and $0.57$~s in our case.

\begin{figure}[tbp]
\includegraphics[width=\grscale\textwidth]{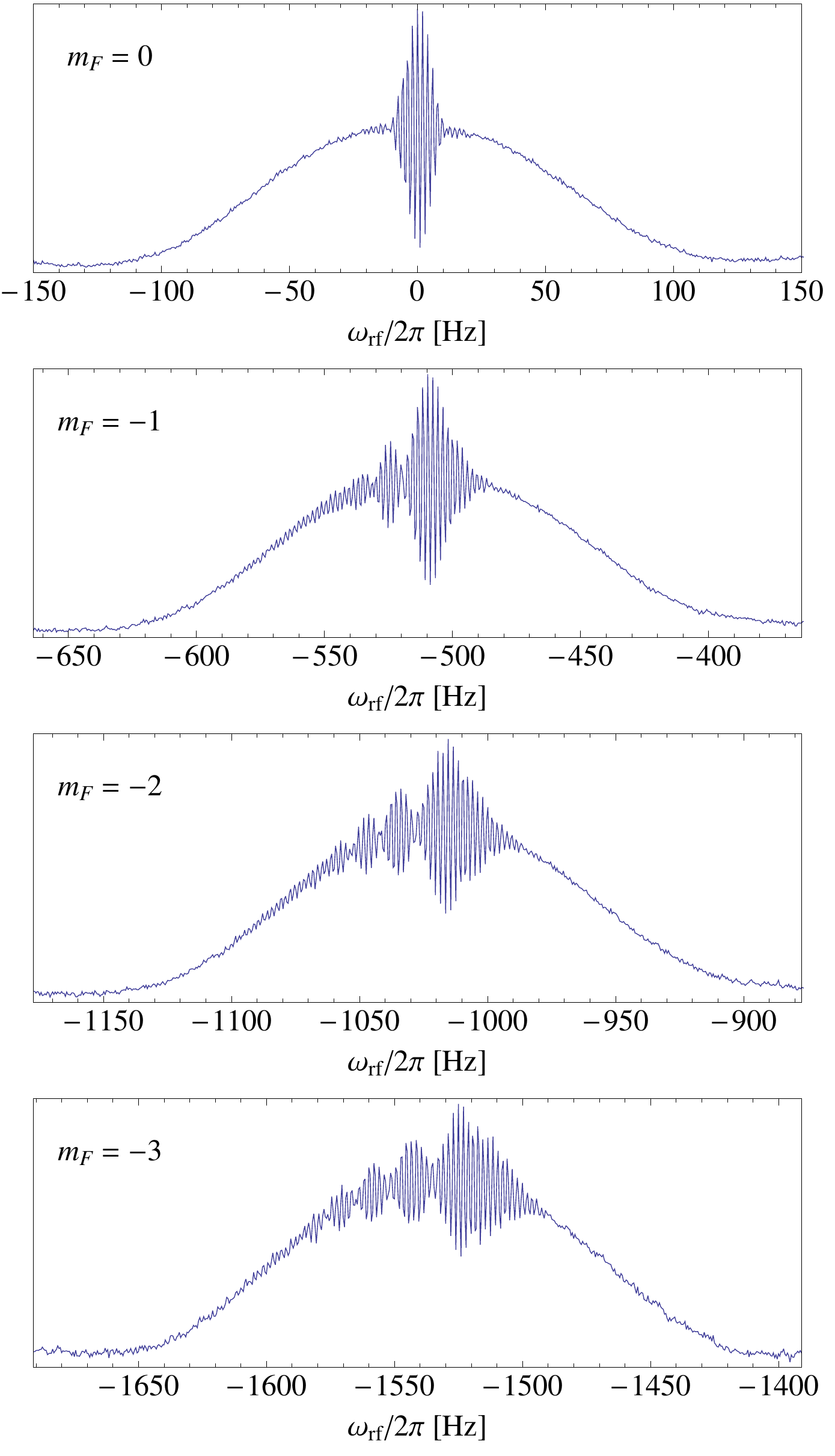}
\caption{Each of these curves, corresponding to $m_{\mathrm{F}}=0,-1,-2,-3$, has been obtained by summing the Ramsey fringes patterns measured for different launching velocities ($3.74$~m/s, $3.80$~m/s, $3.86$~m/s, $3.92$~m/s, $3.98$~m/s, $4.04$~m/s, $4.1$~m/s, $4.16$~m/s, $4.22$~m/s).}
\label{fig:Fig2sum}
\end{figure}

\begin{figure}[tbp]
\includegraphics[width=\grscale\textwidth]{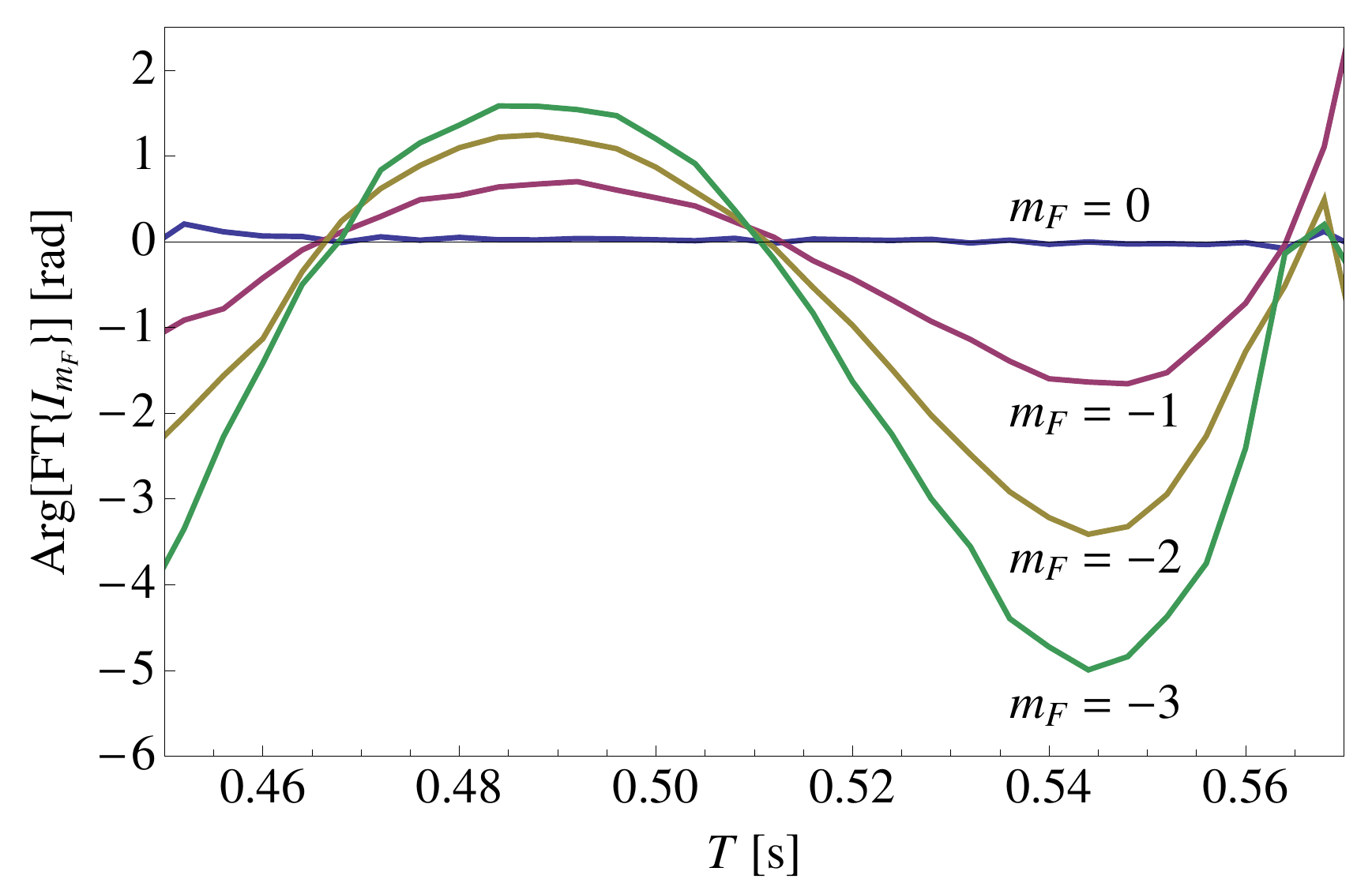}
\caption{Phases of the Fourier transforms measured for $m_\mathrm{F}=0,-1,-2,-3$ as a function of the transit time $T$. Each transit time corresponds to a different height of the apogee above the microwave cavity.}
\label{fig:Fig3}
\end{figure}

To determine the residual phase, and thus the magnetic field spatial variations, it would be useful to find a method allowing to glue together the different phase curves shown in Fig.~\ref{fig:Fig2}. This would be immediate with a measurement of the Ramsey fringes pattern for a very wide atomic beam velocity distribution. Thanks to the linearity of the Fourier transform, one can simulate such a wide velocity distribution by superposing all the Ramsey signals corresponding to the same $m_{\mathrm{F}}$ value but with different launching velocities. The Fourier analysis can be performed on these sums of experimental signals. Indeed, the sum of the different signals are shown in Fig.~\ref{fig:Fig2sum} and the phases obtained from their Fourier transforms in Fig.~\ref{fig:Fig3}. One observes in Fig.~\ref{fig:Fig3} that the phases measured on the different Zeeman components are equal to $-m_\mathrm{F}\varphi_\mathrm{res}(T)$ where $\varphi_\mathrm{res}(T)$ is the phase measured for $m_\mathrm{F}=-1$, as expected from Eq.~(\ref{FTequ}).

%%%%%%%%%%%%%%%%%%%%%%%%%%%%%%%%%%%%%%%%%%%%%%%%%%%%%%%%%%%%%%%%%%%%%%%%%%%%%
\section{Discussion of results}
\label{section5}
%%%%%%%%%%%%%%%%%%%%%%%%%%%%%%%%%%%%%%%%%%%%%%%%%%%%%%%%%%%%%%%%%%%%%%%%%%%%%

%%%%%%%%%%%%%%%%%%%%%%%%%%%%%%%%%%%%%%%%%%%%%%%%%%%%%%%%%%%%%%%%%%%%%%%%%%%%%
%\section{Relation between the residual phase and the time average magnetic field}
%\label{section3}
%%%%%%%%%%%%%%%%%%%%%%%%%%%%%%%%%%%%%%%%%%%%%%%%%%%%%%%%%%%%%%%%%%%%%%%%%%%%%

%A change of $B_0$ implies a change of $\varphi_\mathrm{res}(T)$ according to :
%\begin{eqnarray}
%  B_0 &\rightarrow& B_0'=B_0 + \Delta B \nonumber\\
%  \varphi_\mathrm{res}(T) &\rightarrow& \varphi_\mathrm{res}'(T) = \varphi_\mathrm{res}(T) %-2\pi K_z \Delta B\,T \nonumber
%\end{eqnarray}

%Therefore, one can chose $B_0$ such that the residual phase is zero for the average transit %time:
%\begin{equation}
%	\varphi_\mathrm{res}(\overline{T})=2\pi K_z \int_0^{\overline{T}} B_\mathrm{res}\left(z(t)
%\right)\,dt=0
%\end{equation}
%The direct consequence being that:
%\begin{equation}
%	\overline{B}(\overline{T})=\frac{1}{\overline{T}}\int_0^{\overline{T}} B\left(z(t)\right)
%\,dt=B_0
%\end{equation}

%For other values of $T$, the average magnetic field is given by:
%\begin{equation}
%	\overline{B}(T)=\frac{1}{T}\int_0^{T} B\left(z(t)\right)\,dt=B_0+\frac{\varphi_\mathrm{res}
%(T)}{2\pi K_z\,T}
%\end{equation}

\subsection{In situ magnetometry}
\label{magnetometry}

\begin{figure}[tbp]
\includegraphics[width=\grscale\textwidth]{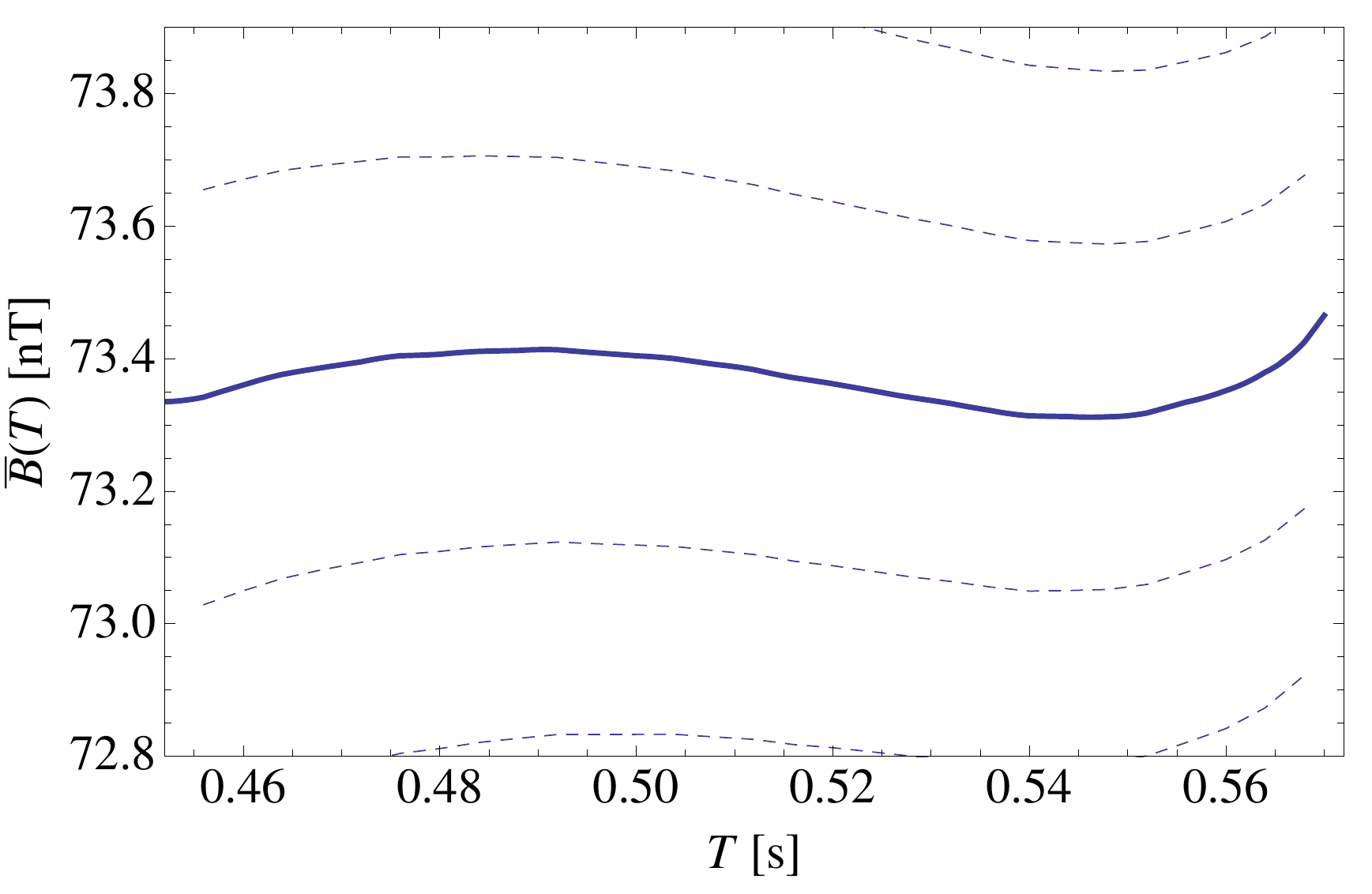}
\caption{Time average of the magnetic field, calculated from the phase of the Fourier transform of the Ramsey fringes measured on $m_\mathrm{F}=-1$. Due to the $2\pi$ ambiguity of the phase, this function is not uniquely determined. The different curves represent possible realizations, they differ by $n/(K_z T)$ where $n$ is an integer. The solid line corresponds to $n=0$.}
\label{fig:Fig6}
\end{figure}

According to our theoretical analysis (section~\ref{section2}), the phase of the Fourier transform obtained for $m_\mathrm{F}=-1$ is equal to $\varphi_\mathrm{res}(T)$ defined by Eq.~(\ref{defphires}). As a consequence, one obtains the time average of the magnetic field along the atomic trajectory $z(t)$ according to :
\begin{eqnarray}
	\overline{B}(T)&=&\frac{1}{T}\int_0^{T} B\left(z(t)\right)\,dt \\
	&=&\frac{1}{T}\int_0^{T} \left(B_0+B_\mathrm{res}(z(t))\right)\,dt \\
	&=&B_0+\frac{\varphi_\mathrm{res}(T)}{2\pi K_z\,T}
\end{eqnarray}
In principle, this last equation allows one to calculate $\overline{B}(T)$ from the Fourier transform of Ramsey fringes. However, $\varphi_\mathrm{res}(T)$ is obtained by calculating the phase of the Fourier transform which inevitably results in a $2\pi$ ambiguity. As a consequence, the average magnetic field may take the following values:
\begin{equation}
	\overline{B}(T)=B_0+\frac{1}{2\pi K_z\,T}\left(\varphi_\mathrm{res}(T)+n\,2\pi\right)
\label{BTcurves}
\end{equation}
where $n$ is any integer number. The resulting graphs are shown in Fig.~\ref{fig:Fig6}, the solid line corresponding to $n = 0$.

This ambiguity in the determination of $\overline{B}(T)$ deserves a few comments. Firstly, by extending the domain of transit time values, it would be possible to distinguish the correct curve ($n=0$) for $\overline{B}(T)$ from the others ($n\neq 0$). Indeed, from Eq.~(\ref{BTcurves}) it is clear that only the $n=0$ curve does not diverge when $T=0$. In pulsed fountains, this ambiguity is solved by throwing the atoms at different altitudes from $T=0$ to its nominal value. However, in our continuous fountain the transit time values are limited by geometrical constraints. Secondly, this $2\pi$ ambiguity in the phase results in a $0.3$~nT ambiguity of $\overline{B}(T)$, which corresponds to a $2$~Hz indetermination on the microwave frequency. This is the distance between two consecutive Ramsey fringes, therefore, determining the correct curve ($n=0$) is a problem equivalent to finding the central fringe in the Ramsey pattern (see section~\ref{centralFringeSection}). In the evaluation process of the continuous fountain FOCS-2, we used a complementary method (time resolved Zeeman spectroscopy) to measure the magnetic field spatial profile and thus lift the above mentioned ambiguity~\cite{EvalReport2}. This technique consists in applying short pulses of an oscillating magnetic field in order to induce Zeeman transitions ($\Delta m_\mathrm{F}=\pm 1$) at different positions along the atomic trajectory. It allowed us to identify the $n=0$ curve in Fig.~\ref{fig:Fig6}. However, its spatial resolution is limited, especially at the apogee where the spread of atomic beam is maximum, and therefore the Fourier analysis method presented in this article provides precious informations about the magnetic field in the region of the apogee.

Finally, in the graph of Fig.~\ref{fig:Fig6} we observe that $\overline{B}(T)$ shows a local minimum $T_a$ and a local maximum $T_b$. When the distribution of transit times is large enough to cover both extrema, the superposition of individual Ramsey signals will be constructive when $T\approx T_a$ and $T\approx T_b$. These two contributions give rise to Ramsey fringes with slightly different periods $1/(2T_a)$ and $1/(2T_b)$, which explains the appearance of beat-like Ramsey patterns in Fig.~\ref{fig:Fig2sum}. However, the complete reconstruction of Ramsey fringes is more complex than that, mainly due to the effect of the magnetic field which produces a $T$ dependent Zeeman shift, and will be discussed in detail in sections~\ref{reconstructionFringesSection} and~\ref{sectionShift}.

\subsection{Reconstruction of Ramsey fringes}
\label{reconstructionFringesSection}

\begin{figure*}[tbp]
\includegraphics[width=\textwidth]{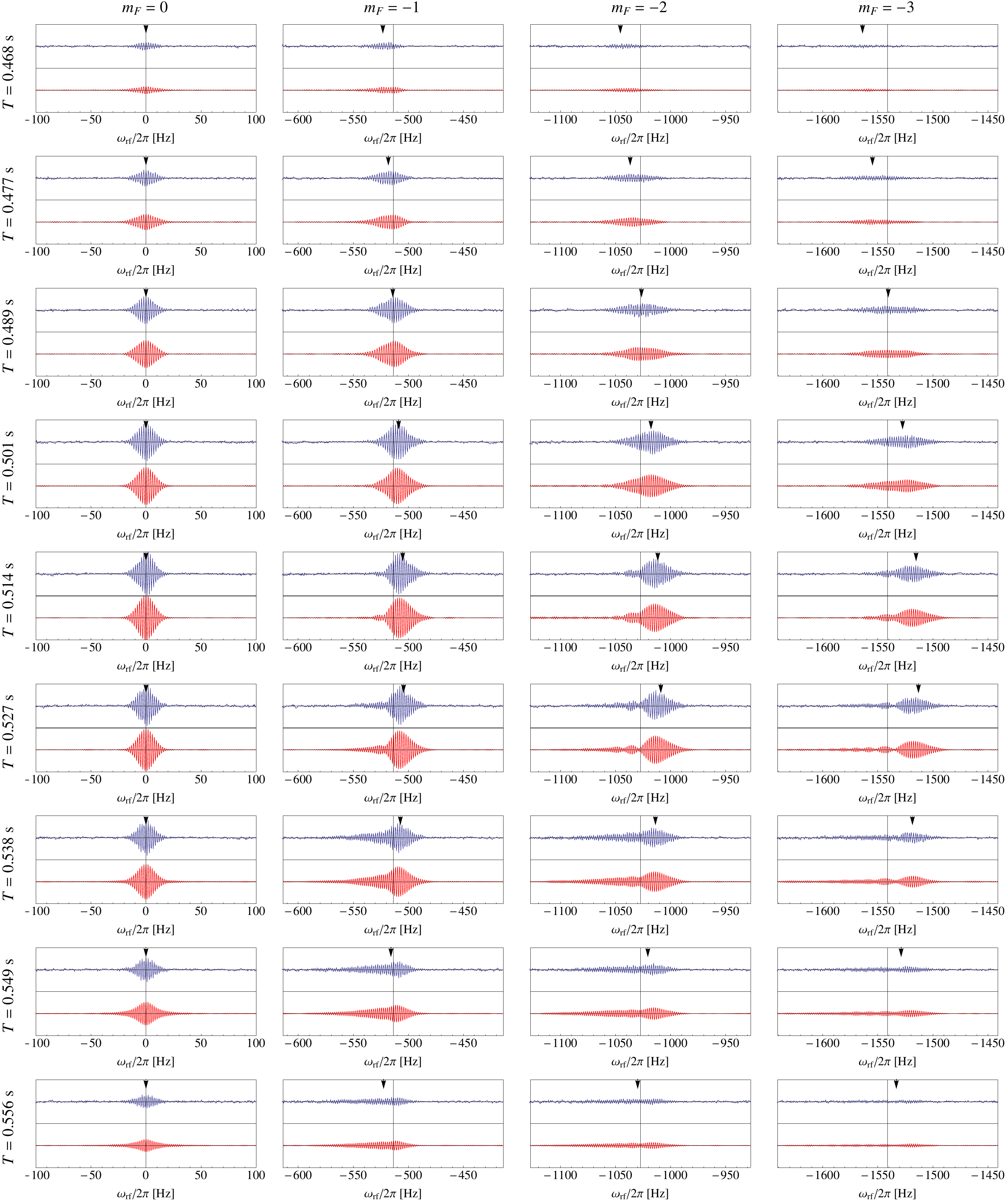}
\caption{Comparison of the measured (upper blue curve) and calculated (lower red curve) Ramsey fringes. The four columns correspond to $m_\mathrm{F}=0,-1,-2,-3$  respectively. The rows correspond to different launching velocities ($3.74$~m/s, $3.80$~m/s, $3.86$~m/s, $3.92$~m/s, $3.98$~m/s, $4.04$~m/s, $4.1$~m/s, $4.16$~m/s, $4.22$~m/s from top to bottom) and thus to different average transit times (indicated on the left side). The calculated fringes are obtained by summing the individual signals as explained in section~\ref{reconstructionFringesSection}. Both the measured and calculated fringes are shown with the Rabi pedestal subtracted. We should emphasize that all the Ramsey fringes are reconstructed using the same time average magnetic field $\overline{B}(T)$. See sections~\ref{reconstructionFringesSection} and~\ref{sectionShift} for details.}
\label{fig:Fig7}
\end{figure*}

With the knowledge of the transit time distributions $\rho(T)$ and of the measured average magnetic field  $\overline{B}(T)$, it is possible to recalculate all the Ramsey fringes by summing the individual Ramsey signals according to:
\begin{equation}
	I_{m_\mathrm{F}}(\omega_\mathrm{rf}) = 
	\frac{1}{2}I_0\int_0^{\infty}\rho(T)\left[ 1 + \cos\left( \varphi_{m_\mathrm{F}}(\omega_\mathrm{rf},T) \right)\right]\,dT 
\label{reconstruction1}
\end{equation}
with:
\begin{equation}
  \varphi_{m_\mathrm{F}}(\omega_\mathrm{rf},T) = 
  (\omega_\mathrm{rf}-\omega_\mathrm{0})T - m_\mathrm{F} 2\pi K_z \overline{B}(T)\,T
\label{reconstruction2}
\end{equation}
The results are presented in Fig.~\ref{fig:Fig7} for every Zeeman components and launching velocities used in the experiment. We should emphasize that all the Ramsey fringes shown in Fig.~\ref{fig:Fig7} are reconstructed using the same function $\overline{B}(T)$ for the time average magnetic field in Eq.~(\ref{reconstruction2}). Considering the fact the frequency sampling interval of the measured Ramsey fringes is $0.5$~Hz, the agreement with the calculated Ramsey fringes is very good. 
%Although one may argue that the good agreement is not surprising since we just calculated the Fourier transform and back, the surprise comes from the fact that the behavior of the individual phases in Eq.~(\ref{reconstruction2}) (for all $m_\mathrm{F}$ values) is described by a unique function $\overline{B}(T)$ which is interpreted as the average magnetic field.

\subsection{Position of the central fringe}
\label{centralFringeSection}

For a given transit time $T$, one can define the position of the central fringe as:
\begin{equation}
  \omega_\mathrm{c} = 
  \omega_\mathrm{0} + m_\mathrm{F} 2\pi K_z \overline{B}(T)
\end{equation}
In other words, it is the microwave frequency for which there is no dephasing between the microwave and the atomic dipole during the free evolution time $T$. Since the transit time is not unique, the same is true for the position of the central fringe. The distribution of central fringe positions is given by $\rho(\omega_\mathrm{c})=\rho(T) dT/d\omega_\mathrm{c}$. In the situation of FOCS-2, the function $\overline{B}(T)$ changes smoothly on the extent of the transit time distribution. Therefore, one can estimate the parameters of the distribution of the central fringes as follows. The average position is given by:
\begin{eqnarray}
  \left\langle \omega_\mathrm{c}\right\rangle &=& 
  \omega_\mathrm{0} + m_\mathrm{F} 2\pi K_z \langle\overline{B}(T)\rangle \\
  &\approx& \omega_\mathrm{0} + m_\mathrm{F} 2\pi K_z \overline{B}(\langle T\rangle)
\label{centralfringepos}
\end{eqnarray}
and the standard deviation by:
\begin{equation}
  \sigma\left( \omega_\mathrm{c}\right) \approx 
  m_\mathrm{F} 2\pi K_z \overline{B}'(\langle T\rangle)\,\sigma(T)
\end{equation}
where $\langle T\rangle$ and $\sigma(T)$ are the average and standard deviation of the transit time distribution $\rho(T)$. Here we should make an important remark: the ambiguity of $\overline{B}(T)$ discussed in section~\ref{magnetometry} results in an ambiguity of the position of the central fringe. Indeed, by inserting Eq.~(\ref{BTcurves}) into Eq.~(\ref{centralfringepos}) one obtains:
\begin{equation}
  \left\langle \omega_\mathrm{c}\right\rangle = 
  \omega_\mathrm{0} + m_\mathrm{F} 2\pi K_z B_\mathrm{0}
  + m_\mathrm{F} \frac{\varphi_\mathrm{res}(\langle T\rangle)}{\langle T\rangle}
  + n \frac{2\pi m_\mathrm{F}}{\langle T\rangle}
\end{equation}
where $n$ is any integer number. However, let's note that this ambiguity does not affect the shape and position of the reconstructed Ramsey fringes. This will be discussed in more detail in the next section.

\subsection{Position of the observed Ramsey pattern}
\label{sectionShift}

Ramsey fringes appear when the individual Ramsey signals interfere constructively. Observing Eq.~(\ref{eqmf}) it is clear that constructive interference can appear only if the phase $\varphi_{m_\mathrm{F}}(\Omega,T)$ exhibits small variations when $\hat{\rho}(T)$ is maximum. Therefore, the position $\omega^{*}_\mathrm{rf}$ of the overall fringe pattern on the frequency axis is given by imposing the condition $\partial \varphi_{m_\mathrm{F}}/\partial T =0$ with $\varphi_{m_\mathrm{F}}$ given in Eq.~(\ref{reconstruction2}):
\begin{equation}
   \frac{\partial \varphi_{m_\mathrm{F}}}{\partial T} =
   \frac{\partial}{\partial T}\left[ (\omega^{*}_\mathrm{rf}-\omega_\mathrm{0}) T -
   m_\mathrm{F} 2\pi K_z \overline{B}(T) T \right]
   = 0
\label{equ22}
\end{equation}
In order to evaluate this condition, we suppose that variations of $\overline{B}(T)$ are small on the extent of the transit time distribution $\rho(T)$. This is only partially fulfilled in our experiment for a given launching velocity, but it is instructive since it helps in understanding the role of $\overline{B}(T)$ in the formation of the fringe pattern. With this assumption, Equ.~(\ref{equ22}) becomes:
\begin{equation}
  \omega^{*}_\mathrm{rf} \approx 
  \omega_\mathrm{0} + m_\mathrm{F} 2\pi K_z \overline{B}(\langle T\rangle)
  + m_\mathrm{F} 2\pi K_z \overline{B}'(\langle T\rangle)\langle T\rangle
\label{fringesPos}
\end{equation}
The first term is the position of the unperturbed atomic transition, then comes the linear Zeeman shift, and the third shift is induced by a first order variation of $\overline{B}(T)$. This expression deserves a few comments. Firstly, the position of the Ramsey pattern given in Eq.~(\ref{fringesPos}) differs from the position of the central fringe given in Eq.~(\ref{centralfringepos}) and the difference is given by the last term proportional to $\overline{B}'(\langle T\rangle)$. Secondly, the linear Zeeman shift, calculated from the measurement of $\overline{B}(T)$ shown in Fig.~\ref{fig:Fig6}, is much too small to explain the shift in position of the Ramsey patterns observed in Fig.~\ref{fig:Fig7}. On the other hand, we calculated the fringe pattern positions $\omega^{*}_\mathrm{rf}$ according to Eq.~(\ref{fringesPos}) and reported them as vertical arrows in each measurement of Fig.~\ref{fig:Fig7}, and we observe that the agreement with the experimental fringes is good. Finally, we should note that adding $1/(K_z T)$ to $\overline{B}(T)$ does not shift the Ramsey pattern since the second and third terms of Eq.~(\ref{fringesPos}) cancel each other. This explains why the position of the Ramsey fringe pattern is not affected by the ambiguity of $\overline{B}(T)$ shown in Eq.~(\ref{BTcurves}).

\subsection{In situ velocimetry}
\label{secVT}

\begin{figure}[tbp]
\includegraphics[width=\grscale\textwidth]{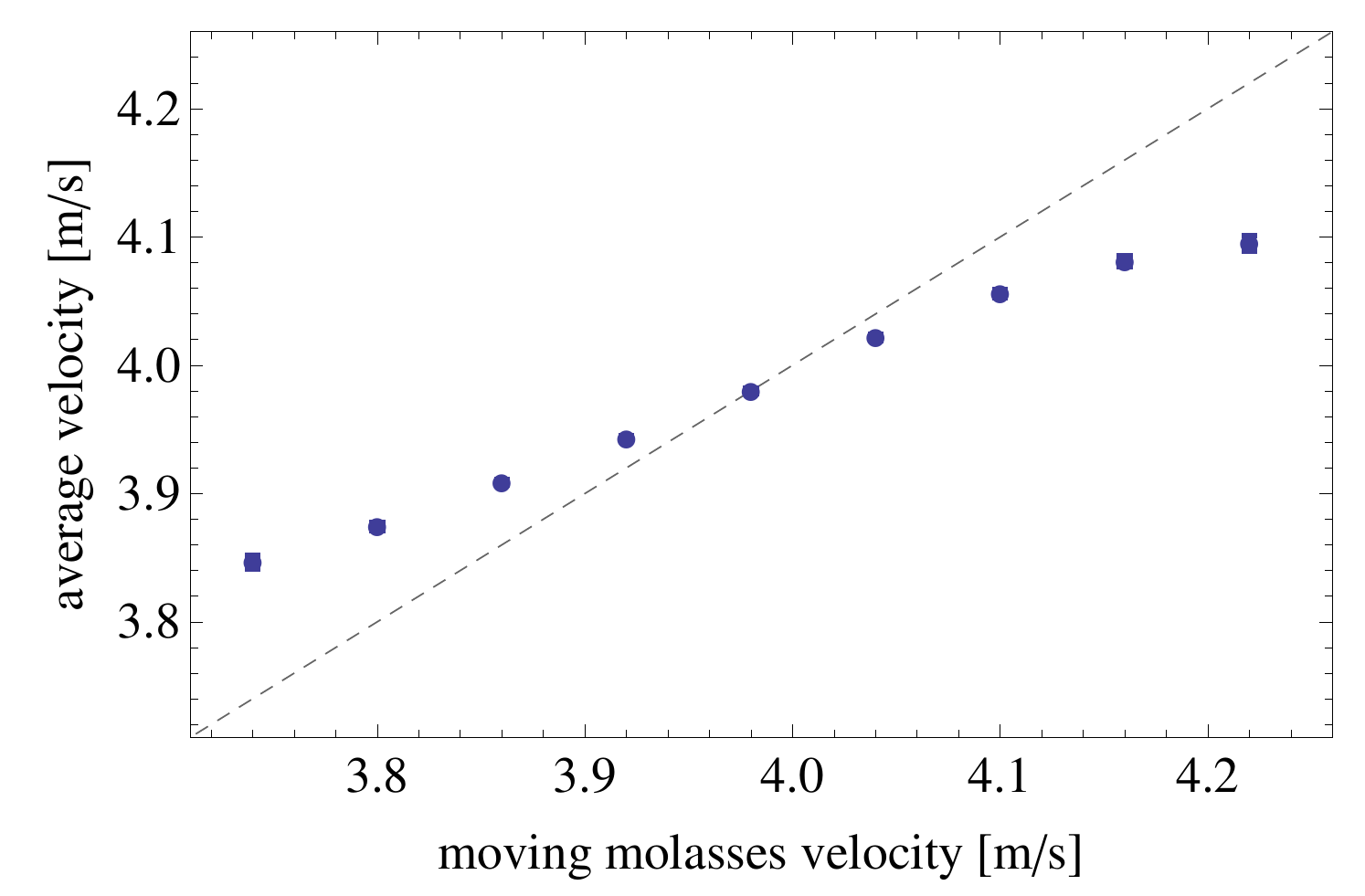}
\includegraphics[width=\grscale\textwidth]{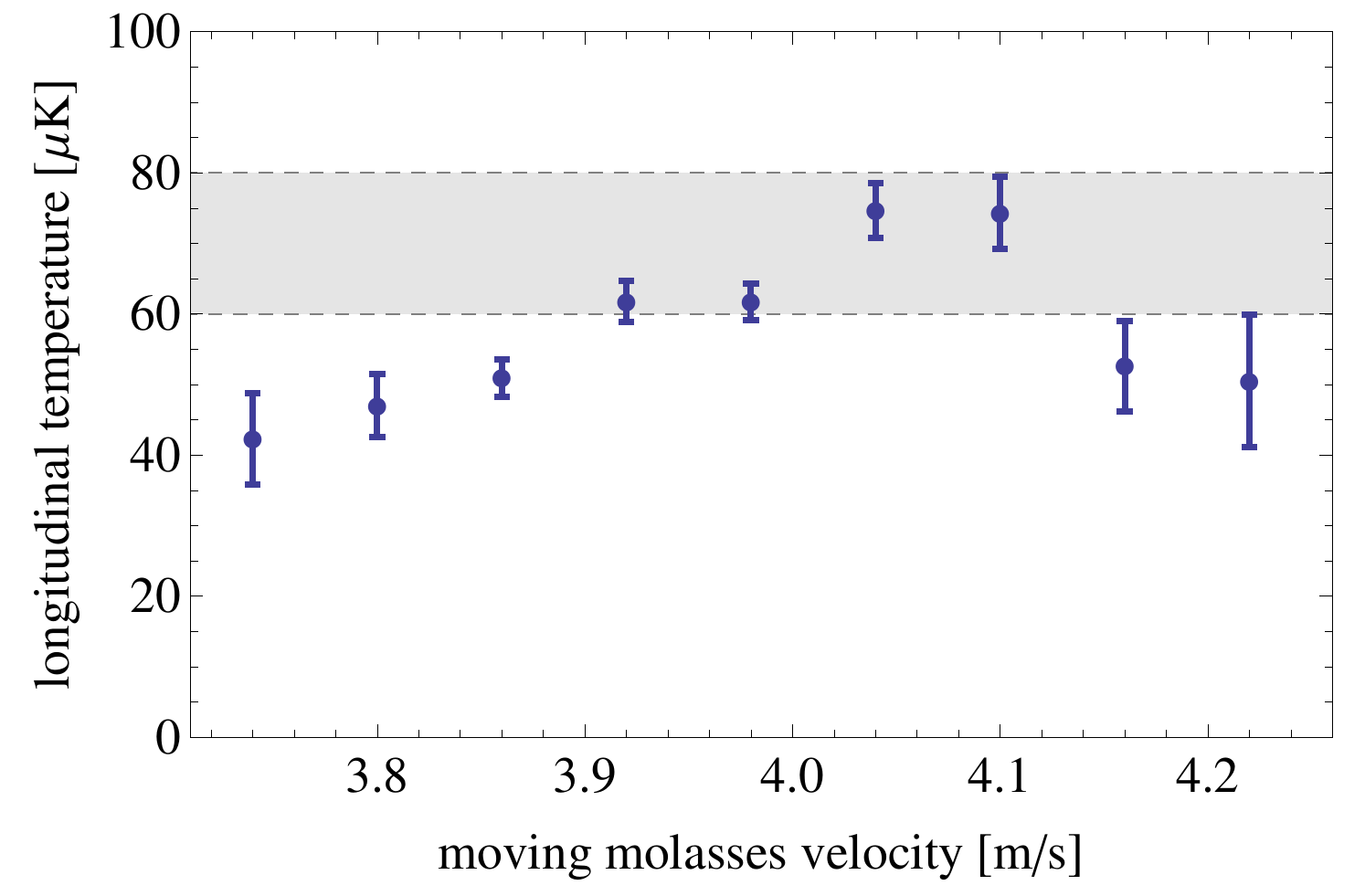}
\caption{Average velocity (upper graph) and longitudinal temperature (lower graph) of the atoms contributing to the Ramsey signal, at the exit of the moving molasses, as a function of the launching velocity. For comparison, the dashed line (upper graph) and the gray band (lower graph) show the values measured in previous experiments using the time-of-flight technique~\cite{PBe99}. See section~\ref{secVT} for details.}
\label{fig:Fig5}
\end{figure}

From the measured distributions of transit times (see Fig.~\ref{fig:Fig2}) we calculated the distributions of velocities at the altitude of the microwave cavity and at the exit of the moving molasses which is situated $48.5$~cm below. Then we used these distributions to calculate the average velocity and longitudinal temperature. The results are displayed in Fig.~\ref{fig:Fig5} for the exit of the moving molasses. 

%\begin{figure}[tbp]
%\includegraphics[width=\grscale\textwidth]{vitesseAvg0}
%\includegraphics[width=\grscale\textwidth]{temperature0}
%\caption{Effective velocity (upper graph) and longitudinal temperature (lower graph) in the %microwave cavity, as a function of the moving molasses launching velocity.}
%\label{fig:Fig4}
%\end{figure}

In order to check the validity of our analysis, we compare the atomic beam velocity and longitudinal temperature at the exit of the moving molasses with the values obtained in previous experiments using the time-of-flight technique (TOF)~\cite{PBe99}. In Fig.~\ref{fig:Fig5}, we observe that the velocity values are in good agreement at the nominal launching velocity of $4.0$~m/s which gives the maximum flux. For other launching velocities, the velocity distribution is truncated by geometrical selection due to diaphragms on the atomic beam trajectory (see Figs.~\ref{fig:Fig1}-\ref{fig:Fig2}) and therefore the values obtained from our analysis of Ramsey fringes do not replicate the actual launching velocity. This is also visible on the longitudinal temperature values which have been measured to be between $60$ and $80$~$\mu$K using the TOF technique. As shown in Fig.~\ref{fig:Fig5}, the temperature values obtained with the Ramsey analysis are indeed compatible with those obtained by TOF for velocities close to $4.0$~m/s. However they decrease for higher or lower velocities, in agreement with the explanation of geometrical selection.

%%%%%%%%%%%%%%%%%%%%%%%%%%%%%%%%%%%%%%%%%%%%%%%%%%%%%%%%%%%%%%%%%%%%%%%%%%%%%
\section{Conclusion}
\label{section6}
%%%%%%%%%%%%%%%%%%%%%%%%%%%%%%%%%%%%%%%%%%%%%%%%%%%%%%%%%%%%%%%%%%%%%%%%%%%%%

We applied Fourier analysis to the Ramsey fringes observed in a continuous atomic fountain clock. By analyzing the Ramsey patterns for every Zeeman component and for different transit times, we have shown that the phase of the Fourier transform is directly linked to the time-averaged magnetic field $\overline{B}(T)$ seen by the atoms during their free evolution of duration $T$. This allowed us to measure $\overline{B}(T)$ over the region of the apogee, with an ambiguity of $n/(K_z T)$ resulting from the $2\pi$ indetermination of the phase. We discussed the role of this ambiguity and showed that it has no influence on the shape and position of the Ramsey pattern. Moreover, this analysis allowed us to establish an expression for the frequency shift of the overall Ramsey pattern induced by spatial variations of the magnetic field. We showed that the position of the Ramsey pattern differs from the position of the so called central fringe by a term proportional to $T \overline{B}'(T)$. In our atomic fountain, the variation of this term induced by a change of transit time $T$ is more important than the corresponding variation of the linear Zeeman shift. Finally, we also showed that the module of the Fourier transform can be interpreted as the distribution of transit times. We used this information to obtain the atomic beam average velocity and longitudinal temperature. The results are in good agreement with previous measurements made with the time-of-flight technique. The method developed in this article, concerning the measurement of the time averaged magnetic field, will be used for the evaluation of the second order Zeeman shift in our continuous atomic fountain FOCS-2. Applicability to atom interferometers is also forseen.

\section*{Acknowledgments}

This work was supported by the Swiss National Science Foundation (grant 200020-121987/1 and Euroquasar project 120418), the Swiss Federal Office of Metrology (METAS), and the University of Neuch\^atel.

%%%%%%%%%%%%%%%%%%%%%%%%%%%%%%%%%%%%%%%%%%%%%%%%%%%%%%%%%%%
%\bibliography{biblioFourierRamsey2}
%%%%%%%%%%%%%%%%%%%%%%%%%%%%%%%%%%%%%%%%%%%%%%%%%%%%%%%%%%%

\end{document}